\documentclass[fleqn,twoside]{article}%
\usepackage{amssymb}
\usepackage{amsmath}
\usepackage{espcrc2}
\usepackage{amsfonts}
\usepackage{graphicx}%
\begin{document}

\title{\textbf{Mass singularity and confinement in QED}$_{3}$}
\author{Yuichi Hoshino\\Kushiro National College of Tehnology,Otanoshike nishi2-32-1,Kushiro City,Hokkaido,084-0916,Japan}

\begin{abstract}
Infrared behaviour of the fermion propagator is examined by spectral
representation.Assuming asymptotic states and using LSZ reduction formula we
evaluate the the lowest order spectral function by definition.After
exponentiation of it we derive the non perturbative propagator.It shows
confinement and dynamical mass generation explicitly.

\end{abstract}
\maketitle

\section{Introduction}

To search the infrared behaviour of the propagator in three dimensional QED
ordinary perturbative method is not appropriate because of its bad infrared
divergences[4].Assuming spectral representation and LSZ reduction formula,we
evaluate the infrared behaviour or the mass singularity of the fermion
propagator in three dimensional QED as in four dimension.In quenched case with
bare photon mass we obtain the Coulomb energy and Dynamical mass by lowest
intermediate state.After exponentiation of this leads the non-perturbative
propagator. Including massless vacuum polarization the short distance part of
the propagator has a logarithmic correction.In the long distance Coulomb force
is screened to suppress mass generation.Confining property and renormalization
constants are discussed.

\section{Position space propagator}

\subsection{\bigskip spectral function}

Here we assume spectral representation of the propagator which is defined by%
\begin{align}
{\normalsize S}_{F}(p) &  {\normalsize =}\int d^{3}x\exp(-ip\cdot
x)\left\langle {\normalsize \Omega|T\psi(x)}\overline{{\normalsize \psi}%
}{\normalsize (0)|\Omega}\right\rangle \nonumber\\
&  \mathbf{=}\int{\normalsize ds}\frac{\gamma\cdot p{\normalsize \rho}%
_{1}(s){\LARGE +{\normalsize \rho}}_{2}(s)}{{\normalsize p}^{2}-s{\LARGE +}%
{\normalsize i\epsilon}},\\
\frac{{\normalsize 1}}{{\normalsize \pi}}{\normalsize \Im S}_{F}(p) &
{\normalsize =}\int d{\normalsize s\delta(p}^{2}-s)[{\normalsize \gamma\cdot
p\rho}_{{\normalsize 1}}{\LARGE ({\normalsize s})+}{\normalsize \rho}%
_{2}{\normalsize (s)]}\nonumber\\
&  {\normalsize =\gamma\cdot p\rho}_{{\normalsize 1}}{\normalsize (p)+\rho
}_{2}{\normalsize (p)}{\LARGE .}%
\end{align}
{\LARGE \ }%
\begin{align}
{\normalsize \rho(p)}{\normalsize =(2\pi)} &  ^{2}\sum_{{\LARGE N}%
}{\normalsize \delta(p-p}_{{\LARGE N}}{\normalsize )}\int{\normalsize d}%
^{{\LARGE 3}}{\normalsize x}\exp{\normalsize (ip\cdot x)}\nonumber\\
&  {\LARGE \times}\left\langle {\normalsize \Omega|\psi(x)|N}\right\rangle
\left\langle {\normalsize N|}\overline{{\normalsize \psi}}%
{\normalsize (0)|\Omega}\right\rangle .
\end{align}
{\normalsize Total three-momentum of the state }$|N\rangle${\normalsize \ is
}$p_{N}^{\mu}.${\normalsize The only intermediates }$N${\normalsize \ contain
one spinor and an arbitrary number of photons.Setting}%
\begin{equation}
|N\rangle=|r;k_{1},...,k_{n}\rangle,
\end{equation}
{\normalsize where }$r${\normalsize \ is the momentum of the spinor }%
$r^{2}=m^{2},${\normalsize and }$k_{i}${\normalsize \ is the momentum of }%
$i${\normalsize th soft photon,we have}%
\begin{align}
\rho(p) &  =\int d^{3}x\exp(-ip\cdot x)\int\frac{md^{2}r}{r^{0}}\nonumber\\
\times\sum_{n=0}^{\infty}\frac{1}{n!} &  (\int\frac{d^{3}k}{(2\pi)^{3}}%
\theta(k_{0})\delta(k^{2})\sum_{\epsilon})_{n}\delta(p-r-\sum_{i=1}^{n}%
k_{i})\nonumber\\
&  \times\left\langle \Omega|\psi(x)|r;k_{1},..,k_{n}\right\rangle
\left\langle r;k_{1},..k_{n}|\overline{\psi}(0)|\Omega\right\rangle .
\end{align}
{\normalsize Here the notations }%
\begin{equation}
(f(k))_{0}=1,(f(k))_{n}=\prod_{i=1}^{n}f(k_{i})
\end{equation}
{\normalsize have been introduced to denote the phase space integral of each
photon.The initial sum over }$\epsilon${\normalsize \ is a sum over
polarization of photon.To evaluate the contribution of the soft-photons,we
consider when only the }$n$ {\normalsize th photon is soft.Our main problem is
the detemination of the matrix element.Here we define the following matrix
element}%
\begin{align}
T_{n} &  =\left\langle \Omega|\psi|r;k_{1},..k_{n}\right\rangle \nonumber\\
&  =\left\langle \Omega|\psi a_{in}^{+}(k_{n})|r;k_{1},..k_{n-1}\right\rangle
.
\end{align}
{\normalsize We consider }$T_{n}${\normalsize \ for }$k_{n}^{2}\neq
0,${\normalsize we continue off the photon mass-shell by
Lehmann-Symanzik-Zimmermann(LSZ)formula:}%
\begin{align}
T_{n} &  =\epsilon_{n}^{\mu}T_{n\mu},\nonumber\\
\epsilon_{n}^{\mu}T_{n}^{\mu} &  =\frac{i}{\sqrt{Z_{3}}}\int d^{3}y\exp
(ik_{n}\cdot y)\nonumber\\
&  \times\square_{y}\left\langle \Omega|T\psi(x)\epsilon^{\mu}A_{\mu
}(y)|r;k_{1},...,k_{n-1}\right\rangle \nonumber\\
&  =-\frac{i}{\sqrt{Z_{3}}}\int d^{3}y\exp(ik_{n}\cdot x)\nonumber\\
&  \times\left\langle \Omega|T\psi(x)\epsilon^{\mu}j_{\mu}(y)|r;k_{1}%
,...,k_{n-1}\right\rangle ,
\end{align}
provided%
\begin{align}
\square_{x}T(\psi A_{\mu}(x) &  =T\psi\square_{x}A_{\mu}(x)\nonumber\\
&  =T\psi(-j_{\mu}(x)+\frac{d-1}{d}\partial_{\mu}^{x}(\partial\cdot A(x)),\\
\partial\cdot A^{(+)}|phys &  >=0,
\end{align}
{\normalsize where the electromagnetic current is}%
\begin{equation}
j^{\mu}(x)=-e\overline{\psi}(x)\gamma_{\mu}\psi(x).
\end{equation}
{\normalsize From the definition (8) }$T_{n}^{\mu}${\normalsize \ is seen to
satisfies Ward-Takahashi-identity:}%
\begin{equation}
k_{n\mu}T_{n}^{\mu}(r,k_{1},..k_{n})=eT_{n-1}(r,k_{1},..k_{n-1}),r^{2}=m^{2},
\end{equation}
{\normalsize provided the equal-time commutation relations}%
\begin{align}
\partial_{\mu}^{x}T(\psi j_{\mu}(x)) &  =-e\psi(x),\nonumber\\
\partial_{\mu}^{x}T(\overline{\psi}j_{\mu}(x)) &  =e\overline{\psi}(x).
\end{align}
{\normalsize In the Bloch-Nordsieck approximation we have most singular
contributions of photons which are emitted from external lines.In perturbation
theory one photon matrix element }$T_{1}$ {\normalsize is given}%
\begin{align}
&  \left\langle in|T(\psi_{in}(x),ie\int d^{3}y\overline{\psi}_{in}%
(y)\gamma_{\mu}\psi_{in}(y)A_{in}^{\mu}(y))|r;k\text{ }in\right\rangle \\
&  =ie\int d^{3}yd^{3}zS_{F}(x-y)\gamma_{\mu}\delta^{(3)}(y-z)\nonumber\\
&  \times\exp(i(k\cdot y+r\cdot z))\epsilon^{\mu}(k,\lambda)U(r,s)\nonumber\\
&  =-ie\frac{(r+k)\cdot\gamma+m}{(r+k)^{2}-m^{2}}\gamma_{\mu}\epsilon^{\mu
}(k,\lambda)\exp(i(k+r)\cdot x)U(r,s),
\end{align}
{\normalsize where }$U(r,s)${\normalsize \ is a four-component free particle
spinor with positive energy.}$U(r,s)${\normalsize \ satisfies the relations}%
\begin{align}
(\gamma\cdot r-m)U(r,s) &  =0,\nonumber\\
\sum_{S}U(r,s)\overline{U}(r,s) &  =\frac{\gamma\cdot r+m}{2m}.
\end{align}
{\normalsize In this case the Ward-Takahasi-identity follows}%
\begin{align}
k_{\mu}T_{1}^{\mu} &  =-ie\frac{1}{\gamma\cdot(r+k)-m}(\gamma\cdot
k)U(r,s)\nonumber\\
&  =-ieU(r,s)=eT_{0}.
\end{align}
{\normalsize For general }$T_{n}${\normalsize \ low-energy theorem determines
the structure of non-singular terms in }$k_{n}.${\normalsize \ Detailed
dicussions are given in ref [1] and non-singular terms are irrelevant for the
single particle singularity in four-dimension.Under the same assumption in
three-dimension we have}%
\begin{equation}
T_{n}|_{k_{n}^{2}=0}=e\frac{\gamma\cdot\epsilon}{\gamma\cdot(r+k_{n}%
)-m}T_{n-1}.
\end{equation}
{\normalsize From this relation the }$n${\normalsize -photon matrix element}%
\begin{equation}
\left\langle \Omega|\psi(x)|r;k_{1},..,k_{n}\right\rangle \left\langle
r;k_{1},..k_{n}|\overline{\psi}(0)|\Omega\right\rangle
\end{equation}
{\normalsize is reduced to the product of lowest-order one-photon matrix
element}%
\begin{equation}
T_{n}\overline{T_{n}}=%
{\displaystyle\prod\limits_{j=1}^{n}}
T_{1}(k_{j})T_{1}^{+}(k_{j})\gamma_{0}.
\end{equation}
{\normalsize In this case the spectral function }$\rho${\normalsize \ in (5)
is given by exponentiation of one-photon matrix element,which yields an
infinite ladder approximation for the propagator.Thus we obtain the spectral
function and the propagator in the followings forms}%
\begin{align}
\rho(p) &  =\int d^{3}x\exp(-ip\cdot x)\int\frac{md^{2}r}{r^{0}}\nonumber\\
&  \times\exp(ir\cdot x)\exp(F),\\
F &  =\sum_{\text{one photon}}\left\langle \Omega|\psi(x)|r;k\right\rangle
\left\langle r;k|\overline{\psi}(0)|\Omega\right\rangle \nonumber\\
&  =\int\frac{d^{2}k}{(2\pi)^{2}}\delta(k^{2})\theta(k_{0})\exp(ik\cdot
x)\sum_{\lambda,S}T_{1}\overline{T_{1}},\\
\sum_{\lambda,S}T_{1}T_{1} &  =e^{2}[\frac{(r+k)\cdot\gamma+m}{(r+k)^{2}%
-m^{2}}\gamma^{\mu}\frac{r\cdot\gamma+m}{2m}\nonumber\\
&  \times\gamma^{\nu}\frac{(r+k)\cdot r+m}{(r+k)^{2}-m^{2}}\Pi_{\mu\nu}].
\end{align}

{\normalsize Here }$\Pi_{\mu\nu}${\normalsize \ is the polarization sum}%
\begin{equation}
\Pi_{\mu\nu}=\sum_{\lambda}\epsilon_{\mu}(k,\lambda)\epsilon_{\nu}%
(k,\lambda)=-g_{\mu\nu}-(d-1)\frac{k_{\mu}k_{\nu}}{k^{2}},
\end{equation}
{\normalsize and the free photon propagator is}%
\begin{equation}
D_{0}^{\mu\nu}=\frac{1}{k^{2}+i\epsilon}[g_{\mu\nu}+(d-1)\frac{k_{\mu}k_{\nu}%
}{k^{2}}].
\end{equation}
{\normalsize We get}%
\begin{align}
F  &  =-e^{2}\int\frac{d^{3}k}{(2\pi)^{2}}\exp(ik\cdot x)\theta(k^{0}%
)\nonumber\\
&  \times\lbrack\delta(k^{2})(\frac{m^{2}}{(r\cdot k)^{2}}+\frac{1}{(r\cdot
k)})\nonumber\\
&  +(d-1)\frac{\delta(k^{2})}{k^{2}}].
\end{align}
{\normalsize The second term }$\delta(k^{2})/k^{2}${\normalsize \ equals to
}$-\delta^{\prime}(k^{2}).${\normalsize Our calculation is the same with the
imaginary part of the photon propagator.To avoid infrared divergence which
arises in the phase space integral we must introduce small photon mass }$\mu
${\normalsize \ as an infrared cut-off.Therefore (22) is modified to}%
\begin{align}
F  &  =-e^{2}\int\frac{d^{3}k}{(2\pi)^{2}}\exp(ik\cdot x)\theta(k^{0}%
)\nonumber\\
&  \times\lbrack\delta(k^{2}-\mu^{2})(\frac{m^{2}}{(r\cdot k)^{2}}+\frac
{1}{(r\cdot k)})\nonumber\\
&  -(d-1)\frac{\partial}{\partial k^{2}}\delta(k^{2}-\mu^{2})].
\end{align}
{\normalsize It is helpful to use function }$D_{+}(x)${\normalsize \ to
determine }$F${\normalsize \ }%
\begin{align}
D_{+}(x)  &  =\frac{1}{(2\pi)^{2}i}\int\exp(ik\cdot x)d^{3}k\theta
(k^{0})\delta(k^{2}-\mu^{2})\nonumber\\
&  =\frac{1}{(2\pi)^{2}i}\int_{0}^{\infty}J_{0}(kx)\frac{\pi kdk}{2\sqrt
{k^{2}+\mu^{2}}}=\frac{\exp(-\mu x)}{8\pi ix}.
\end{align}
{\normalsize If we use parameter trick(exponential cut-off) }%
\begin{align}
\lim_{\epsilon\rightarrow0}\int_{0}^{\infty}d\alpha\exp(i(k+i\epsilon
)\cdot(x+\alpha r))  &  =i\frac{\exp(ik\cdot x)}{k\cdot r},\\
\lim_{\epsilon\rightarrow0}\int_{0}^{\infty}\alpha d\alpha\exp(i(k+i\epsilon
)\cdot(x+\alpha r))  &  =-\frac{\exp(ik\cdot x)}{(k\cdot r)^{2}},
\end{align}

\begin{align}
F_{1}  &  =ie^{2}m^{2}\int_{0}^{\infty}d\alpha\alpha D_{+}(x+\alpha
r)\nonumber\\
&  =\frac{e^{2}}{8\pi}(\frac{\exp(-\mu\left\vert x\right\vert )}{\mu
}+\left\vert x\right\vert \operatorname{Ei}(\mu\left\vert x\right\vert )),\\
F_{2}  &  =-ie^{2}\int_{0}^{\infty}d\alpha D_{+}(x+\alpha r)\nonumber\\
&  =-\frac{e^{2}}{8\pi m}\operatorname{Ei}(\mu\left\vert x\right\vert ),\\
F_{g}  &  =ie^{2}(d-1)\frac{\partial}{\partial\mu^{2}}D_{+}(x)\nonumber\\
&  =\frac{(d-1)e^{2}}{8\pi\mu}\exp(-\mu\left\vert x\right\vert ).
\end{align}
{\normalsize where the function }$\operatorname{Ei}(\mu\left\vert x\right\vert
)${\normalsize \ is defined }%
\begin{equation}
\operatorname{Ei}(\mu\left\vert x\right\vert )=\int_{1}^{\infty}\frac
{\exp(-\mu\left\vert x\right\vert t)}{t}dt.
\end{equation}
{\normalsize It is understood that all terms which vanishes with }%
$\mu\rightarrow0${\normalsize \ are ignored.The leading non trivial
contributions to }$F${\normalsize \ are }%
\begin{equation}
\operatorname{Ei}(\mu\left\vert x\right\vert )=-\gamma-\ln(\mu\left\vert
x\right\vert )+O(\mu\left\vert x\right\vert ),
\end{equation}

\begin{align}
F_{1}  &  =\frac{e^{2}m^{2}}{8\pi r^{2}}\left(  -\frac{1}{\mu}+\left\vert
x\right\vert (1-\ln(\mu\left\vert x\right\vert )-\gamma)\right)
+O(\mu),\nonumber\\
F_{2}  &  =\frac{e^{2}}{8\pi\sqrt{r^{2}}}(\ln(\mu\left\vert x\right\vert
)+\gamma)+O(\mu),\nonumber\\
F_{g}  &  =\frac{e^{2}}{8\pi}(\frac{1}{\mu}-\left\vert x\right\vert
)(d-1)+O(\mu),
\end{align}%
\begin{align}
F  &  =\frac{e^{2}}{8\pi\mu}(d-2)+\frac{\gamma e^{2}}{8\pi\sqrt{r^{2}}%
}\nonumber\\
&  +\frac{e^{2}}{8\pi\sqrt{r^{2}}}\ln(\mu\left\vert x\right\vert )\nonumber\\
&  -\frac{e^{2}}{8\pi}\left\vert x\right\vert \ln(\mu\left\vert x\right\vert
)-\frac{e^{2}}{8\pi}\left\vert x\right\vert (d-2+\gamma),
\end{align}
{\normalsize where }$\gamma${\normalsize \ is Euler's constant.We set }%
$r^{2}=m^{2}${\normalsize \ in the phase space integral; }%
\begin{align}
\rho(p)  &  =F.T\int\frac{md^{2}r}{\sqrt{r^{2}+m^{2}}}\exp(ir\cdot
x)\exp(F(m,\left\vert x\right\vert ))\nonumber\\
&  =F.T\frac{\exp(-m\left\vert x\right\vert )}{4\pi\left\vert x\right\vert
}\exp(F(m,\left\vert x\right\vert )),\\
\overline{\rho(x)}  &  =\frac{\exp(-m\left\vert x\right\vert )}{4\pi\left\vert
x\right\vert }\exp(F(m,\left\vert x\right\vert )).
\end{align}
{\normalsize Here }$\exp(-m\left\vert x\right\vert )/4\pi\left\vert
x\right\vert ,\left\vert x\right\vert =\sqrt{-x^{2}}${\normalsize \ is a free
scalar propagator with physical mass }$m${\normalsize \ and }$\exp
(F)${\normalsize \ denotes the quantum correction for the propagator involving
infinite numbers of photons in our approximation.}

\subsection{{\protect\normalsize Confining property}}

{\normalsize \bigskip Here we mention the confining property of the propagator
}$S_{F}(x)${\normalsize \ in position space}%
\begin{equation}
S_{F}(x)=(\frac{i\gamma\cdot\partial}{m}+1)[\frac{m\exp(-m\left\vert
x\right\vert )}{4\pi\left\vert x\right\vert }(\mu\left\vert x\right\vert
)^{D(1-m\left\vert x\right\vert )}].
\end{equation}
{\normalsize The }$S_{F}${\normalsize \ dumps strongly at large }%
$x${\normalsize \ provided }%
\begin{equation}
\lim_{x\rightarrow\infty}(\mu\left\vert x\right\vert )^{-Dm\left\vert
x\right\vert }=0.
\end{equation}
{\normalsize The profiles of the }$\overline{\rho(x)}${\normalsize \ for
various values of }$D\geq1${\normalsize \ are shown in Fig.1.The effect of
}$(\mu\left\vert x\right\vert )^{-Dm\left\vert x\right\vert }$%
{\normalsize \ in position space is seen to decrease the value of the
propagator at low energy and shown in Fig.2.}%

\begin{figure}
[ptb]
\begin{center}
\includegraphics[
height=2.2952in,
width=2.2952in
]%
{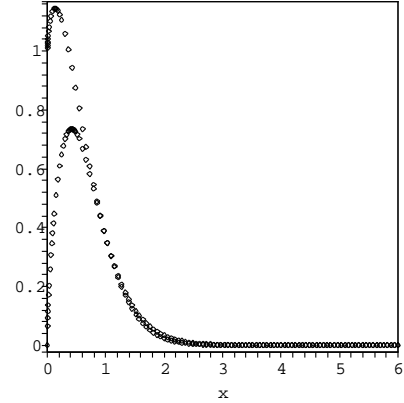}%
\caption{$\overline{\rho(x)}$ for $D=1,1.5,m=\mu=unit$}%
\label{f1}%
\end{center}
\end{figure}
\begin{figure}
[ptb]
\begin{center}
\includegraphics[
height=2.1274in,
width=2.1274in
]%
{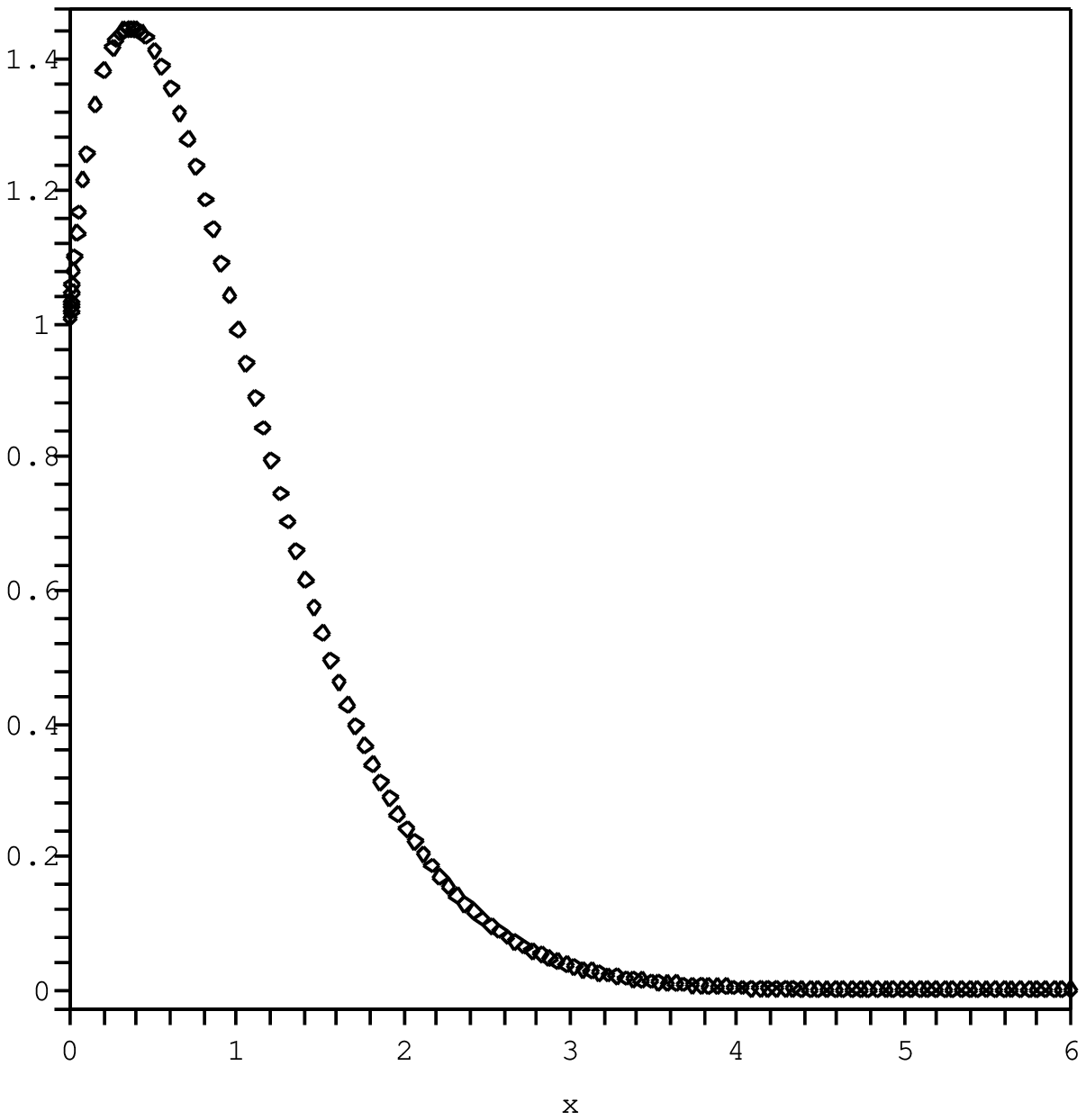}%
\caption{$\left(  \mu\left\vert x\right\vert \right)  ^{-Dm\left\vert
x\right\vert },D=1,m=\mu=unit$}%
\label{f2}%
\end{center}
\end{figure}
{\normalsize In section IV and V we discuss dynamical mass,the renormalization
constant,and bare mass in connection of each terms in }$F${\normalsize .}

{\normalsize After angular integration of (38),we get the propagator}%

\begin{align}
S_{F}(p)  &  =(\frac{\gamma\cdot p}{m}+1)\rho(p),\\
\rho(p)  &  =\frac{m}{2\pi\sqrt{-p^{2}}}\int_{0}^{\infty}d\left\vert
x\right\vert \sin(\sqrt{p^{2}x^{2}})\nonumber\\
&  \times\exp(A-(m_{0}+B)\left\vert x\right\vert )(\mu\left\vert x\right\vert
)^{-C\left\vert x\right\vert +D},
\end{align}
{\normalsize where }%
\begin{align}
A  &  =\frac{e^{2}}{8\pi\mu}(d-2)+\frac{\gamma e^{2}}{8\pi m},B=\frac{e^{2}%
}{8\pi}(d-2+\gamma),\nonumber\\
C  &  =\frac{e^{2}}{8\pi},D=\frac{e^{2}}{8\pi m}.
\end{align}
{\normalsize If we discuss the Euclidean or off-shell propagator we can omitt
the linear infrared divergent part in }$A.${\normalsize In this case }%
$m${\normalsize \ denotes a physical mass }%
\begin{equation}
m=|m_{0}+\frac{e^{2}}{8\pi}(d-2+\gamma)|.
\end{equation}

\section{Analysis in momentum space}

{\normalsize To search the infrared behaviour we expand the propagator in the
powers of the coupling constant }$e^{2}${\normalsize \ and obtained the
Fourier transform of }$\overline{\rho}(x)${\normalsize \ [2,3].In that case it
is not enough to see the structure of infrared behaviour which can be compared
to the well-known four dimensional QED[7].Instead we make Laplace
transformation of }$(\mu\left\vert x\right\vert )^{-Dm\left\vert x\right\vert
},${\normalsize which }{\normalsize leads the general spectral representation
of the propagator in momentum space.After that we show the roles of Coulomb
energy and position dependent mass.The former determines the dimension of the
propagator and the latter acts to change mass.Let us begin to study the effect
of position dependent mass(Self-energy),Coulomb energy in momentum space}%
\begin{align}
\exp(-\left\vert x\right\vert M(x))  &  =\exp(-\left\vert x\right\vert
\frac{e^{2}}{8\pi}\ln(\mu\left\vert x\right\vert ),\\
\exp(-Coulomb\text{ }energy/m)  &  =\exp(-\frac{e^{2}}{8\pi m}\ln
(\mu\left\vert x\right\vert )).
\end{align}
{\normalsize Similar discussion was given to study the effects of self-energy
and bare potential in the stabilty of massless }$e^{+}e^{-}$%
{\normalsize composite in lattice simulation[8].The position space free
propagator }%
\begin{equation}
S_{F}(x,m_{0})=-(i\gamma\cdot\partial+m_{0})\frac{\exp(-m_{0}\left\vert
x\right\vert )}{4\pi\left\vert x\right\vert }%
\end{equation}
{\normalsize is modified by these two terms which are related to dynamical
mass and wave function renormalization.To see this let us think about position
space propagator}%
\begin{equation}
\overline{\rho(x)}=\frac{\exp(-m\left\vert x\right\vert )}{4\pi\left\vert
x\right\vert }(\mu\left\vert x\right\vert )^{-C\left\vert x\right\vert }%
(\mu\left\vert x\right\vert )^{D}.
\end{equation}
{\normalsize It is easy to see that the probability of particles which are
separated with each other in the long distance is suppressed by the factor
}$(\mu\left\vert x\right\vert )^{-C\left\vert x\right\vert },${\normalsize and
the Coulomb energy modifies the short distance behaviour from the bare
}$1/\left\vert x\right\vert ${\normalsize \ to }$1/\left\vert x\right\vert
^{1-D}${\normalsize .The effect of Coulomb energy for the infrared behaviour
of the free particle with mass }$m${\normalsize \ can be seen by its fourier
transform[2,9] }%
\begin{align}
&  4\pi\int_{0}^{\infty}x^{2}\frac{\sin(\sqrt{p^{2}x^{2}})}{\sqrt{p^{2}x^{2}}%
}\frac{\exp(-m\left\vert x\right\vert )}{4\pi\left\vert x\right\vert }%
(\mu\left\vert x\right\vert )^{D}d\left\vert x\right\vert \nonumber\\
&  =\mu^{D}\frac{\Gamma(D+1)\sin((D+1)\arctan(\sqrt{-p^{2}}/m))}{\sqrt{-p^{2}%
}(-p^{2}+m^{2})^{(1+D)/2}}\\
&  \sim\mu^{D}(\sqrt{-p^{2}}-m)^{-1-D}\text{ near }p^{2}=-m^{2}.
\end{align}
{\normalsize Above formula shows the structure in momentum space is modified
for both infrared and ultraviolet regions.Usually constant }$D$%
{\normalsize \ represents the coefficent of the leading infrared divergence
for fixed mass in four dimension.Therefore Coulomb energy in three dimension
has the same effects as in four dimension but change the ultraviolet behaviour
since the coupling constant }$e^{2}${\normalsize \ is not renormalized.Now we
consider the role of }$M(x)${\normalsize \ as the dynamical mass at low
momentum.First we define Fourier transform of the scalar part of the
propagator; }%
\begin{equation}
\rho(p)=F.T(\frac{m\exp(-m\left\vert x\right\vert )}{4\pi\left\vert
x\right\vert }(\mu\left\vert x\right\vert )^{-C\left\vert x\right\vert +D}).
\end{equation}
{\normalsize \ If we use Laplace transformation }%
\[
F(s)=\int_{0}^{\infty}\exp(-s\left\vert x\right\vert )(m\left\vert
x\right\vert )^{-C\left\vert x\right\vert }d\left\vert x\right\vert
\]
{\normalsize we easily see that }$(m\left\vert x\right\vert )^{-C\left\vert
x\right\vert }${\normalsize \ acts as mass changing operator }$m\rightarrow
m-s$%
\begin{align}
&  \exp(-m\left\vert x\right\vert )(m\left\vert x\right\vert )^{-C\left\vert
x\right\vert }\nonumber\\
&  =\int dsF(s)\exp(-(m-s)\left\vert x\right\vert ).
\end{align}
{\normalsize To separate the }$\mu${\normalsize \ dependence we define
}$m^{\ast}=m(1+D\ln(\mu/m)).F(s)$ {\normalsize is shown in Fig.4.}%

\begin{figure}
[ptb]
\begin{center}
\includegraphics[
height=2.1274in,
width=2.1274in
]%
{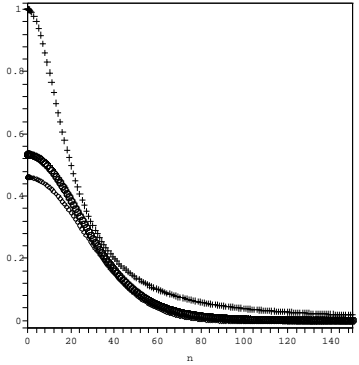}%
\caption{$\rho(p)$ for $D=0(u),1(m),1.5(l),p=n/2$}%
\label{f3}%
\end{center}
\end{figure}
\begin{figure}
[ptb]
\begin{center}
\includegraphics[
height=2.1274in,
width=2.1274in
]%
{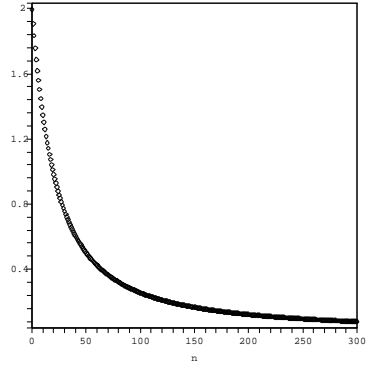}%
\caption{$F(s)$ for $m=\mu=unit,s=n/20$}%
\label{f4}%
\end{center}
\end{figure}

{\normalsize \bigskip}

{\normalsize We have the complete the expression of the spectral function}%
\begin{equation}
\overline{\rho(x)}=\frac{m\mu^{D}\exp(-m^{\ast}\left\vert x\right\vert )}%
{4\pi\left\vert x\right\vert ^{1-D}}\int_{0}^{\infty}\exp(s\left\vert
x\right\vert )F(s)ds,
\end{equation}%
\begin{align}
&  \rho(p)\nonumber\\
&  =m\mu^{D}\Gamma(D+1)\int_{0}^{\infty}dsF(s)\nonumber\\
&  \times\frac{\sin((D+1)\arctan(\sqrt{-p^{2}}/(m-s)))}{\sqrt{-p^{2}}%
(-p^{2}+(m-s)^{2})^{(1+D)/2}}%
\end{align}

\section{{\protect\normalsize Bare mass and vacuum expectation value
}$\left\langle \overline{\psi}\psi\right\rangle $}

{\normalsize In this section we examine the renormalization constant and bare
mass and study the condition of vanishing bare mass based on spectral
representation.The spinor propagator in position space is expressed in the
following for }$\rho_{1}=\rho_{2}=\rho$ {\normalsize which is the case in our
model[2].The equation for the renormalization constant in terms of the
spectral functions read}%

\begin{align}
m_{0}Z_{2}^{-1}  &  =m\int\omega\rho_{2}(\omega)d\omega=\frac{1}{4}%
\lim_{p\rightarrow\infty}tr[p^{2}S_{F}(p)],\\
Z_{2}^{-1}  &  =\int\rho_{1}(\omega)d\omega=\frac{1}{4}\lim_{p\rightarrow
\infty}tr[\gamma\cdot pS_{F}(p)].
\end{align}
{\normalsize we obtain }%
\begin{align}
m_{0}Z_{2}^{-1}  &  \sim\lim_{p^{2}\rightarrow\infty}\sqrt{-p^{2}}%
^{-D}=\left[
\begin{array}
[c]{cc}%
0 & (0<D)\\
m & (0=D)
\end{array}
\right]  ,\\
Z_{2}^{-1}  &  =\lim_{p^{2}\rightarrow\infty}\sqrt{-p^{2}}^{-D}=\left[
\begin{array}
[c]{cc}%
0 & (0<D)\\
1 & (0=D)
\end{array}
\right]  .
\end{align}
{\normalsize This means that propagator in the high energy limit has no part
which is proportional to the free one.Usually mass is a parameter which
appears in the Lagrangean.For example chiral symmetry is defined for the bare
quantity.In ref[6] the relation between bare mass and renormalized mass of the
fermion propagator in QED is discussed based on renormalization group equation
with the assumption of ultraviolet stagnant point and shown that the bare mass
vanishes in the high energy limit even if we start from the finite bare mass
in the theory.It suggests that symmetry properties can be discussed in terms
of renormalized quantities.In QCD bare mass vanishes in the short distance by
asymptotic freedom.And the dynamical mass vanishes too[5].In our apprximation
this problem is understood that at short distance propagator in position space
tends to }%
\begin{equation}
\overline{\rho(x)}_{x\rightarrow0}\rightarrow\left\vert x\right\vert
^{D-1}(\mu\left\vert x\right\vert )^{-C\left\vert x\right\vert },
\end{equation}
{\normalsize where we have }$\overline{\rho(0)}=finite$ {\normalsize at }$D=1$
{\normalsize case which is independent of the bare mass }$m_{0}$%
{\normalsize .Thus we have a same effect as vanishing bare mass in four
dimensional model.Of course we have a dynamical mass generation which is
}$m=|\frac{e^{2}}{8\pi}(d-2+\gamma)+M(x)|$ {\normalsize for }$m_{0}=0$
{\normalsize in our approximation}$.${\normalsize There is a chiral symmetry
at short distance where the bare or dynamical mass vanishes in momentum space
but its breaking must be discussed in terms of the values of the order
parameter.Therefore it is interesting to study the possibility of pair
condensation in our approximation.The vacuum expectation value of pair
condensate is evaluated }%
\begin{align}
\left\langle \overline{\psi}\psi\right\rangle  &  =-trS_{F}(x)=-2\int
_{0}^{\infty}\frac{p^{2}d\sqrt{p^{2}}}{2\pi^{2}}\frac{\Gamma(D+1)\mu^{D}%
}{(D+1)2\sqrt{p^{2}}}\nonumber\\
&  \times\int_{0}^{\infty}dsG(s)\frac{\sin((D+1)\arctan(\sqrt{p^{2}}%
/(m-s))}{\sqrt{((m-s)^{2}+p^{2})^{D+1}}},
\end{align}
{\normalsize is finite for }$D\geq1$ {\normalsize for finite cut-off }$\mu
.${\normalsize In the weak coupling limit we obtain }$Z_{2}=1,m_{0}=m$
{\normalsize and }$\left\langle \overline{\psi}\psi\right\rangle =\infty
${\normalsize .If we introduce chiral symmetry as a global }$U(2n)$%
{\normalsize ,it breaks dynamically into }$SU(n)\times SU(n)\times U(1)\times
U(1)$ {\normalsize as in QCD[9,11] for }$D=1$ {\normalsize for finite infrared
cut-off.Our model may be applicable to relativistic model of super fluidity in
three dimension.Usually we do not find the critical coupling }$D=1$
{\normalsize in the analysis of the Dyson-Schwinger equation in momentum space
where only continuum contributions are taken into account and we do not define
physical mass.Finally we notice the effects of vacuum
polarization[4,10,11,12].In the presence of vacuum polarization with
}${\normalsize N}$ massless fermion {\normalsize the mass shift and dynamical
mass we find in the quenched case are decreased by screening at short
distance.In the long distance mass shitft is strongly suppresd and we have
only wave renormalization as in four dimension.We have an infrared behaviour
of the propagator in the Landau gauge }%
\begin{equation}
S_{F}(p)\simeq(\frac{\mu}{m})^{D^{\prime}}\frac{\gamma\cdot p+m}{2m^{2}%
}(1-\frac{p^{2}}{m^{2}})^{-(1+D^{\prime})},D^{\prime}=\frac{8}{N\pi^{2}}.
\end{equation}

\section{{\protect\normalsize References}}

{\normalsize \noindent\lbrack1]R.Jackiw,L.Soloviev,Phys.Rev.\textbf{173}%
(1968)1458;}

{\normalsize L.D.Landau and E.M.Lifzhits,Quantum Electrodynamics,Pergamon
Press,Oxford(1982);}

{\normalsize C.Itzykson,J-BZuber,QuantumField Theory,McGRAW-HILL.\newline%
[2]Y.Hoshino,\textbf{JHEP}09(2004)048.\newline%
[3]Conrand.J.Burden,Justin.Praschika,Craig.D.Roberts,}

{\normalsize Phys.Rev.\textbf{D46}(1992)2695;ibid,Phys.Rev.\textbf{D47}%
(1993)5581; A.Bashir,A.Raya,Phys.Rev.\textbf{D64}(2001)105001.\newline%
[4]A.B.Waites,R.Delbourgo,Int.J.Mod.Physics.\textbf{A7}(1992)6857.}%
\newline{\normalsize [5]H.D.Politzer,Nucl.Phys.\textbf{B117}(1976)397.\newline%
[6]K.Nishijima,Prog.Theor.Phys.81(1989)878;83(1990)1200.\newline%
[7]L.S.Brown,QuantumFieldTheory,CambridegeUniversityPres(1992);}%
\newline%
[8]{\normalsize E.Dagotto,J.B.Kogut,A.Kocic,Phys.Rev.Lett.62(1988)1083.}%
\newline[9]{\normalsize M.Koopman,DynamicalMassGenerationinQED}$_{3}%
,${\normalsize Ph.D thesis,Groningen University(1990),Chapter4;}

{\normalsize Y.Hoshino,Il.Nouvo.Cim.\textbf{112A}(1999)335.\newline%
[10]YuichiHoshino,preprint,hep-th/0506045.}\newline[11]T.Appelquist,D.Nash,L.C.R.Wijewardhana,

Phys.Rev.Lett.\textbf{60}{\normalsize (1988)2575.}\newline%
[12]K.-I.Kondo,H.Nakatani,Prog.Theor.Phys.87:193-206(1992).\newline

\end{document}